\documentclass[conference]{IEEEtran}
\IEEEoverridecommandlockouts
\usepackage{cite}
\usepackage{amsmath,amssymb,amsfonts}
\usepackage{algorithmic}
\usepackage{graphicx}
\usepackage{textcomp}
\usepackage{xcolor}
\usepackage{wrapfig}
\usepackage[export]{adjustbox}
\usepackage{multicol}
\usepackage{gensymb}
\usepackage{steinmetz}
\usepackage{mathtools}
\usepackage{subfig}
\def\BibTeX{{\rm B\kern-.05em{\sc i\kern-.025em b}\kern-.08em
    T\kern-.1667em\lower.7ex\hbox{E}\kern-.125emX}}
\begin{document}

\title{AI-driven, Model-Free Current Control: A Deep Symbolic Approach for Optimal Induction Machine Performance\\

}

\author{\IEEEauthorblockN{Muhammad Usama}
\IEEEauthorblockA{\textit{Department of Electrical Engineering} \\
\textit{Chosun University}\\
Gwangju, South Korea \\}
\and
\IEEEauthorblockN{Yunkyung Hwang}
\IEEEauthorblockA{\textit{Department of Electrical Engineering} \\
\textit{POSTECH}\\
Pohang, South Korea \\
}
\and
\IEEEauthorblockN{Jaehong Kim}
\IEEEauthorblockA{\textit{Department of Electrical Engineering} \\
\textit{Chosun University}\\
Gwangju, South Korea \\
}
}

\maketitle

\begin{abstract}
This paper proposed a straightforward and efficient current control solution for induction machines employing deep symbolic regression (DSR). The proposed DSR-based control design offers a simple yet highly effective approach by creating an optimal control model through training and fitting, resulting in an analytical dynamic numerical expression that characterizes the data. Notably, this approach not only produces an understandable model but also demonstrates the capacity to extrapolate and estimate data points outside its training dataset, showcasing its adaptability and resilience. In contrast to conventional state-of-the-art proportional-integral (PI) current controllers, which heavily rely on specific system models, the proposed DSR-based approach stands out for its model independence. Simulation and experimental tests validate its effectiveness, highlighting its superior extrapolation capabilities compared to conventional methods. These findings pave the way for the integration of deep learning methods in power conversion applications, promising improved performance and adaptability in the control of induction machines. The simulation and experimental test results are provided with a 3.7kw induction machine to verify the efficacy of the proposed control solution.
\end{abstract}

\begin{IEEEkeywords}
Analytical expression; Current control; LSTM; Deep Symbolic Regression; Induction Machine
\end{IEEEkeywords}

\section{Introduction}
Electric motors, specifically induction motors, have held and continue to hold a fundamental position in industrial sectors. Induction motors have been the preferred option in most industrial applications for many years, primarily due to their high reliability, simplicity, and cost-effectiveness \cite{b1}. Utilizing vector control techniques for induction motor operation allows us to achieve independent control of torque and flux components, resembling the control precision of separately excited DC motors. However, recent research highlights a limitation in conventional vector control strategies, specifically in their current control loop. This limitation impedes the full decoupling of d- and q-axis control, resulting in a lack of complete independence between these control aspects \cite{p1}. Consequently, the system's performance becomes more susceptible to disturbances, such as variations in system parameters and external load perturbations \cite{p2,aa5}.

The effective control of current in induction machines is critical for guaranteeing maximum performance and energy efficacy in various industrial applications. While substantial advances have been achieved in speed control loops using artificial intelligence (AI), the use of AI-based solutions in current control is still largely unexplored. Unlike speed control, where AI approaches like as neural networks and fuzzy logic have been intensively studied and deployed \cite{aa1,aa2}, current control in induction machines has received very little attention. Despite the essential requirement of current control in providing exact torque regulation and dynamic response in induction machines, standard control approaches have limits when dealing with nonlinearities, uncertainties, and disturbances observed in real-world industrial environments. As a result, there is a critical need for intelligent control techniques capable of handling these challenges while providing optimal control performance.

Over the past few years, substantial research efforts have focused on utilizing artificial intelligence to optimize nonlinear systems, yielding promising results. However, a notable drawback in using such networks is their inherent complexity and low explainability. Neural network architectures are often likened to black boxes, challenging to interpret, and heavily reliant on gradient-based search methods. Therefore, there is a demand for an understandable model that can extract meaningful insights from intricate datasets and generalize its applicability beyond the initial training data. In this context, exclusively data-driven techniques can effectively adapt to the dynamics of the system \cite{a1}.

The main goal of this paper is to introduce nonlinear control law for optimal current control in induction machines, ensuring robust performance under both standstill and transient conditions. It leverages deep symbolic optimization, combining symbolic regression with deep learning, to develop an analytical model that replaces conventional PI-based current control models. The objective is to derive a numerical expression that best fits the dataset, utilizing a large neural network (NN) to create a concise and interpretable mathematical equation, eliminating the need for NN interpretation. Additionally, this control scheme relies solely on the current errors, making it entirely independent of a particular plant model, making it resilient against adversarial disturbances. This work aims to pave the road for the development and deployment of new AI-driven current control techniques, which can improve the performance of induction machines in varied industrial applications.

\begin{figure*}
	\centerline{\includegraphics [width=14cm]{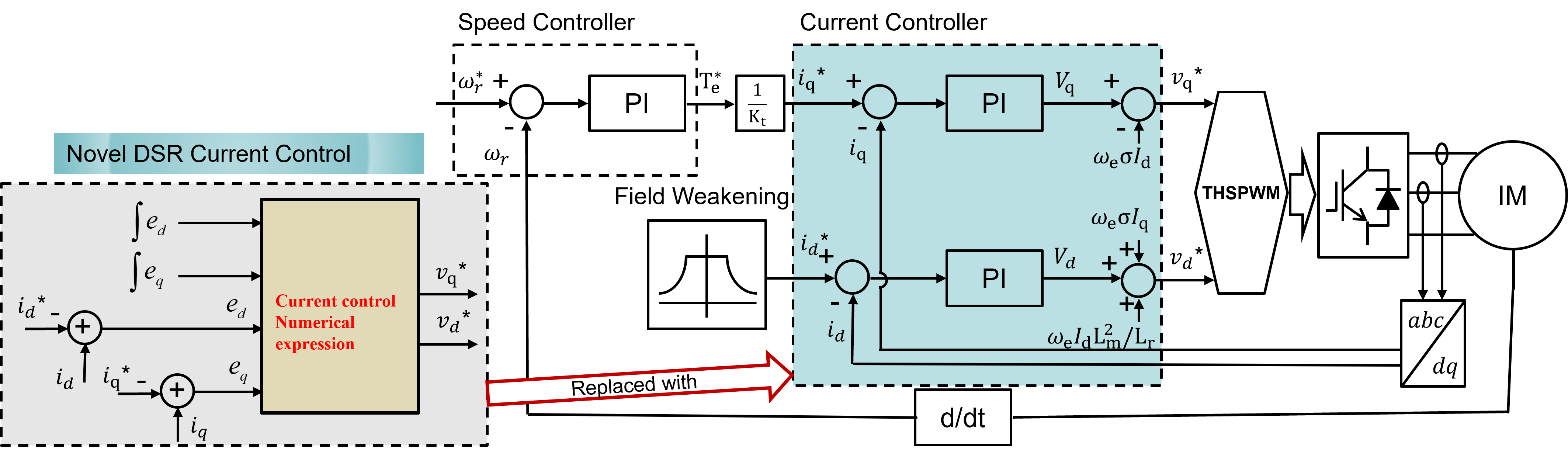}}
	\caption{Proposed DSR vector control for an induction machine.}
	\label{fig1}
\end{figure*}

\section{Mathematical Model of Induction Machine in d-q Frame}
The electromagnetic behaviour of an induction motor in the d-q reference frame, which rotates synchronously, can be described as follows\cite{b2}: 

\begin{multline}
\frac{d}{d t}\left[\begin{array}{c}
i^{e}_{d s} \\
i^{e}_{q s} \\
\lambda^{e}_{d r} \\
\lambda^{e}_{q r}
\end{array}\right]=\\
\left[\begin{array}{cccc}
-\left(\frac{R_s}{L_\sigma}+\frac{R_r L_m^2}{L_r^2 L_\sigma}\right) & \omega_e & \frac{R_r L_m}{L_r^2 L_\sigma} & \frac{\omega_r L_m}{L_r L_o} \\
-\omega_e & -\left(\frac{R_s}{L_\sigma}+\frac{R_r L_m^2}{L_r^2 L_\sigma}\right) & -\frac{\omega_r L_m}{L_r L_r} & \frac{R_r L_m}{L_r^2 L_\sigma} \\
\end{array}\right.\\
\left.\begin{array}{cccc}
\frac{R_r L_m}{L_r} & 0 & -\frac{R_r}{L_r} & \left(\omega_{\mathrm{e}}-\omega_r\right) \\
0 & \frac{R_r L_m}{L_r} & -\left(\omega_e-\omega_r\right) & -\frac{R_r}{L_r}
\end{array}\right] \\
\left[\begin{array}{c}
i^{e}_{d s} \\
i^{e}_{q s} \\
\lambda^{e}_{d r} \\
\lambda^{e}_{q r}
\end{array}\right]+\frac{1}{L_\sigma}\left[\begin{array}{c}
V^{e}_{d s} \\
V^{e}_{q s} \\
0 \\
0
\end{array}\right]
\label{eq1}
\end{multline}

where d-axis and q-axis stator currents in the synchronous reference frame are denoted by $i^{e}_{d s}$ and $i^{e}_{q s}$; the d-axis and q-axis rotor flux linkages are denoted by $\lambda^{e}_{d r}$ and $\lambda^{e}_{q r}$ respectively; $V^{e}_{d s}$ and $V^{e}_{q s}$ signify the d-axis and q-axis stator voltages; ${R_s}$ and ${R_r}$ denote the stator and rotor resistances; ${L_s}$ and ${L_r}$ stands for stator and rotor inductances; ${L_m}$ and ${L_\sigma}$ stand for mutual and leakage inductances; $\omega_e$ and $\omega_r$ are the electrical angular speed and rotor angular speed, respectively.

Utilizing rotor field-oriented scheme for current control design, where $\lambda_{qr}^{e}=0$ then the stator currents are derived as: 

\begin{equation}
\begin{aligned}
\frac{d i^{e}_{d s}}{d t}= & -\left(\frac{R_s}{L_\sigma}+\frac{R_r L_m^2}{L_r^2 L_\sigma}\right) i^{e}_{d s}+\omega_e i_{q s}+\frac{R_r L_m}{L_r^2 L_\sigma} \lambda_{d r} 
& +\frac{1}{L_\sigma} V^{e}_{d s} \\
\frac{d i^{e}_{q s}}{d t}= & -\left(\frac{R_s}{L_\sigma}+\frac{R_r L_m^2}{L_r^2 L_\sigma}\right) i^{e}_{q s}-\omega_e i_{d s}-\frac{\omega_r L_m}{L_r L_\sigma} \lambda_{d r} 
& +\frac{1}{L_\sigma} V^{e}_{q s}
\end{aligned}
\label{eq2}
\end{equation}

The torque produced by the induction machine is given as follow:

\begin{equation}
T_e = \frac{3P}{4}\frac{L_m}{L_r}\lambda^{e}_{dr}i^{e}_{qs}.
\label{eq3}
\end{equation}

The traditional vector control scheme employs a cascaded PI control structure with an inner control loop having a larger bandwidth than the outer control loop to ensure excellent drive performance. The larger bandwidth with the compensation terms will effectively decouple the motor electrical dynamics from mechanical dynamics which will help to enhance stability and provide robustness. However, unlike traditional control, the proposed AI-driven control scheme can effectively achieve separate and precise control over both torque and flux as shown in Figure \ref{fig1}. We use a hierarchical control technique to precisely regulate torque and flux in induction machines. The outer loop has a Proportional-Integral (PI) controller, which ensures steady speed control by altering the speed reference signal based on motor speed error. The inner loop has an AI-based current control design that uses machine learning to dynamically alter current references in real-time. Unlike traditional approaches that rely on mathematical models, an AI-driven solution optimises current references using neural networks to ensure optimal torque and flux characteristics.
 
\begin{figure}[htbp]
	\centerline{\includegraphics [width=10cm]{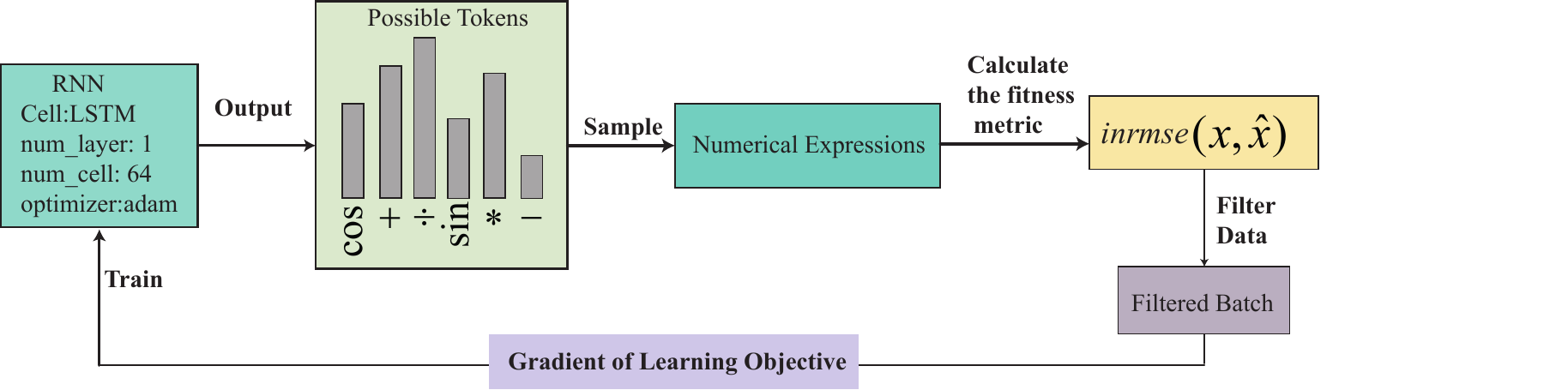}}
	\caption{DSR architecture.}
	\label{fig2}
\end{figure}

\begin{figure}[htbp]
	\centerline{\includegraphics [width=9cm]{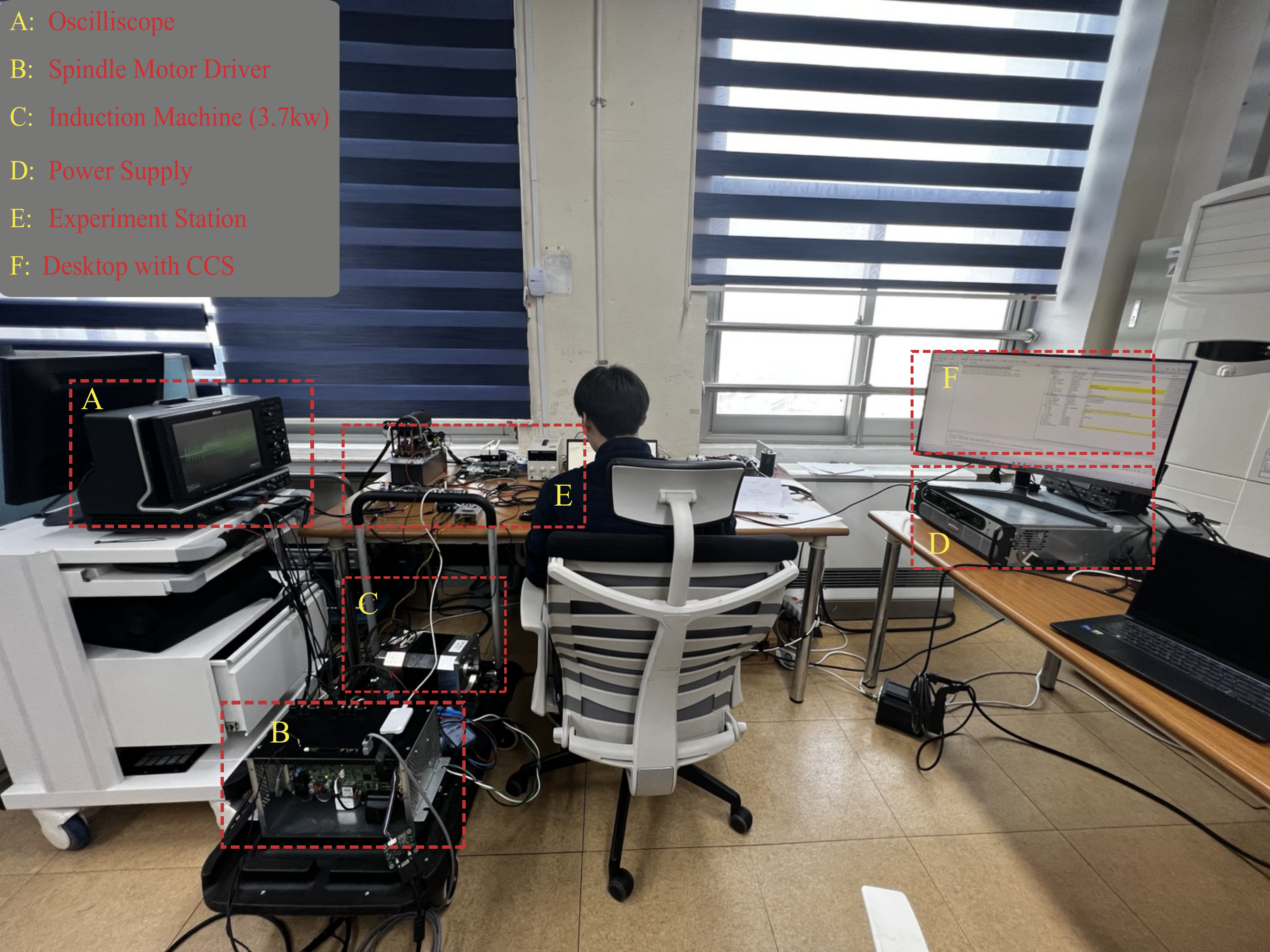}}
	\caption{Experimental Setup.}
	\label{fig4}
\end{figure}

\section{Proposed AI-driven Current Control}

Artificial intelligence applications in electric motor drives have garnered significant attention in research in recent years. In particular, neural networks are becoming more and more well-known because of their exceptional performance and quick dynamic reaction in comparison to conventional control systems. A variety of identification techniques, such as local linear regression, reinforcement learning, and fuzzy logic, are presented in the literature to improve motor drive performance. Artificial intelligence (AI) has been used recently to tackle more difficult problems, such as speed sensorless control based on artificial neural networks (ANNs), torque observers powered by ANNs, and motor parameter identification driven by AI \cite{aa3,aa4}. In addition, neural networks' ability to estimate complex functions makes them a desirable control method for motor drives, as it does not require exact mathematical models.

For industrial applications, conventional AI-based control schemes face challenges because neural networks are complex and often seen as black box. They're hard to understand, and interpret, and rely heavily on specific data patterns found through gradient searches. Additionally, the long-standing problem of finding analytical expressions to effectively fit complex data persists. To address these challenges, we introduce Deep Symbolic Regression (DSR) \cite{a1}, an innovative deep learning approach that utilizes the capabilities of deep neural networks to create models that are easy to understand and implement in real systems. DSR-based control method consists of converting mathematical expressions into sequences, creating an autoregressive model capable of generating expressions within predefined constraints and implementing a risk-seeking policy gradient approach to train the model in generating more accurate expressions. The training is conducted offline using Python, employing inverse normalized root-mean-squared error (INRMSE) as a cost function. The block diagram of the proposed control solution is illustrated in Figure \ref{fig2}. 

The proposed current controller's operating methodology is detailed as follows: Current errors and their integration in the rotating reference frame are evaluated at a sampling time interval indicated as $T_s$;
The system incorporates the integrated error information, ensuring that there are no steady-state errors in reference tracking; The DSR algorithm formulates numerical expressions that are both understandable and well-suited to the dataset utilising a risk-seeking policy gradient approach;
 Finally, these generated expressions are subsequently used within an online model to serve as an optimal current controller

Following thorough training on a diverse dataset acquired through simulations, the final equations that optimized the reward and minimised the cost function for the reference $v_d^*$ and $v_q^*$ were obtained as follows:

\begin{equation}
\begin{aligned}
v_{d}^* &= 13 \cdot x1 - \sin(x1-x4), \\
v_{q}^* &= 12 \cdot x2 + x3 + 2 \cdot x4 \\
&\quad+ (x1^2 + x1 - x2 - x4) \cdot \sin(x1) \\
&\quad+ \sin(x1 \cdot (-x1 + 2 \cdot x3) - x2) \\
&\quad+ \cos(2 \cdot x2).
\end{aligned}
\label{eq4}
\end{equation}

where $x1=\Delta{i_{ds}},x2=\Delta{i_{qs}}, x3 = \int{\Delta{i_{ds}}}, ~and~~ x4 = \int{\Delta{i_{qs}}}$ respectively. Finally, employing these expressions the obtained reference voltage $(V_{dq}^*)$ will be fed to the voltage source inverter.

\begin{table}[htbp]
	\caption{Experimental Parameters}
	\begin{center}
		\begin{tabular}{|c|c|}
			\hline
			Parameters & Values\\
			\hline
			\hline
			Stator Resistance ($R_s$) & 0.7025 ${\Omega}$ \\
			\hline
			Rotor Resistance ($R_r$) & 0.8783 ${\Omega}$\\
			\hline
			Stator Inductance ($L_s$) & 0.032 H\\
			\hline
			Rotor Inductance ($L_r$) & 0.032 H\\
			\hline
			Mutual Inductance ($L_m$) & 0.031 H\\
			\hline
			DC Voltage ($V_{dc}$)  & 311 v\\
			\hline
			Inertia ($J$) &  0.006 $kg m^2$\\
			\hline
			Pole pair($P$) & 2\\
			\hline
			Switching Frequency ($F_{s}$) & 16k Hz\\
			\hline
					\end{tabular}
		\label{tab1}
	\end{center}
\end{table}

\subsection{Test Results}


\begin{figure*}
\centering
\subfloat[PI current controller.] {\includegraphics[width=9cm]{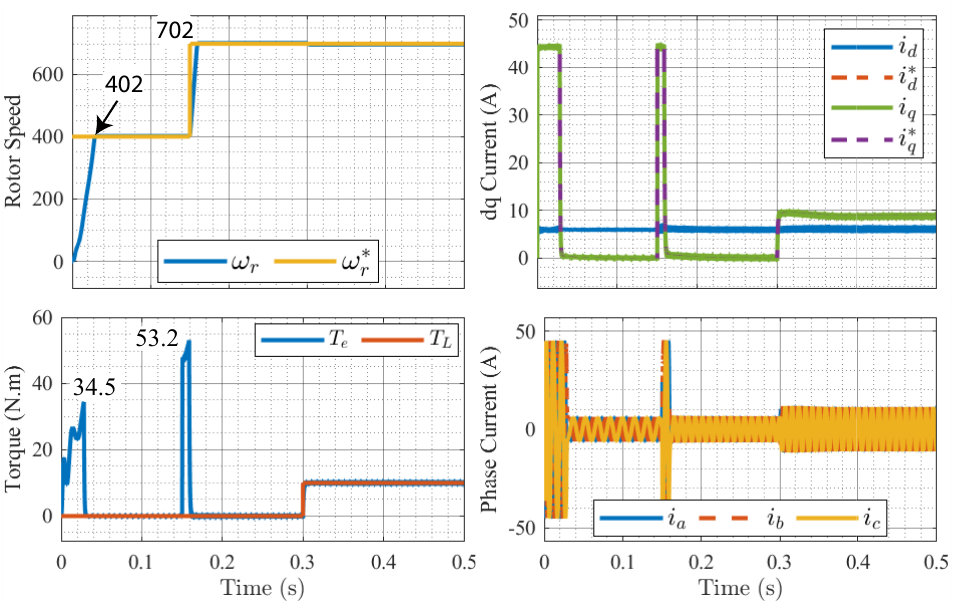} \label{mot_pi}} 
\subfloat[DSR-based current controller.]{\includegraphics[width=9cm]{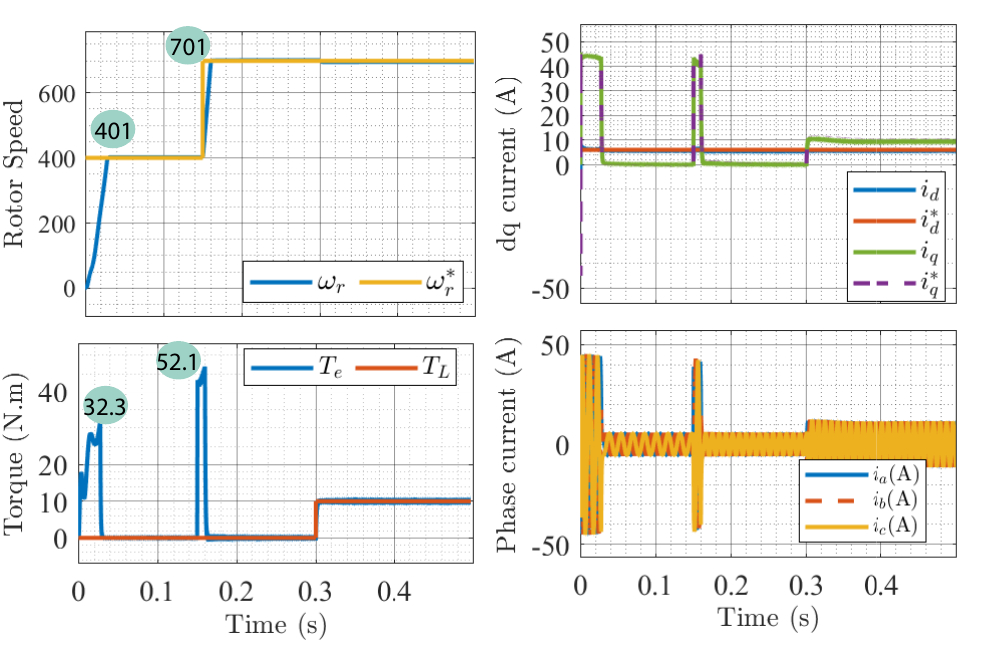} \label{mot_dsr}}
\caption{Simulation results.}
\label{fig5}
\end{figure*}

\begin{figure}
\centering
\subfloat[At 500rpm.] {\includegraphics[width=9cm]{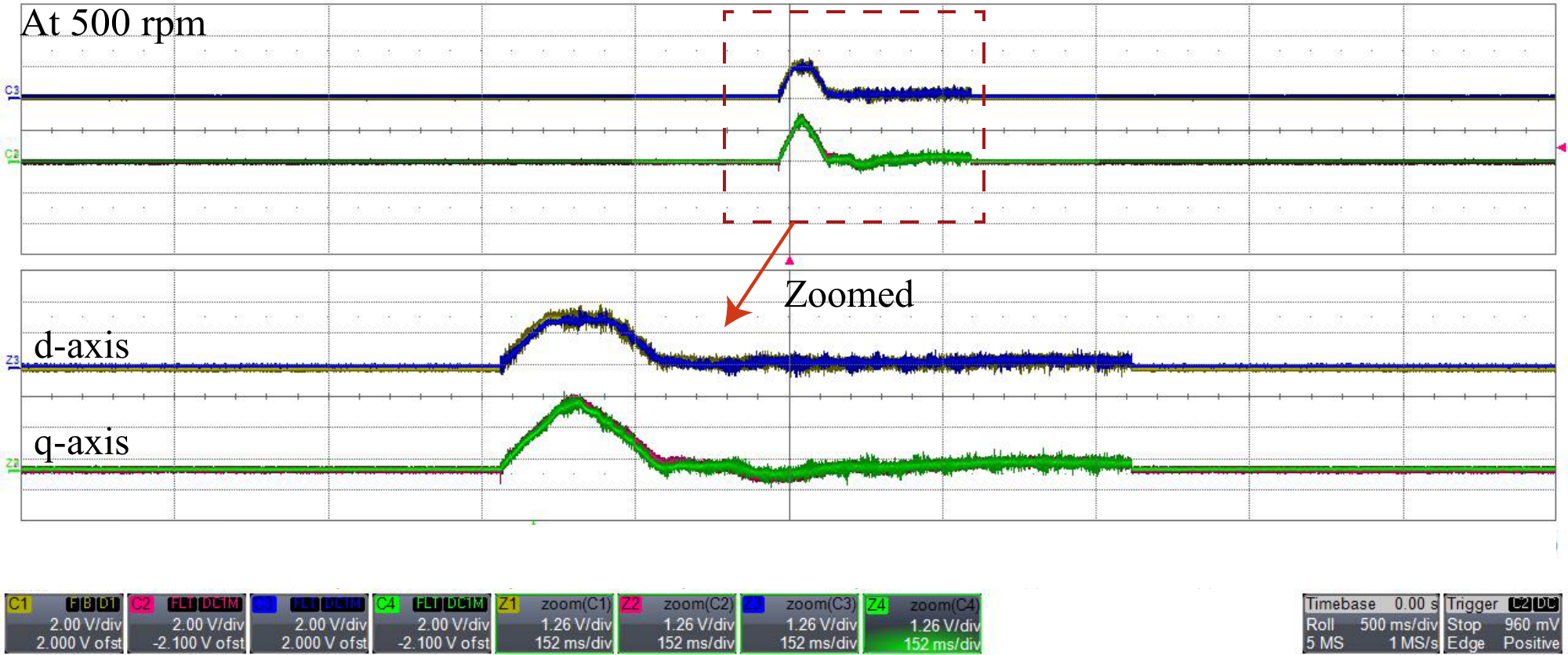} \label{fig6a}} \\
\subfloat[At 2000rpm.]{\includegraphics[width=9cm]{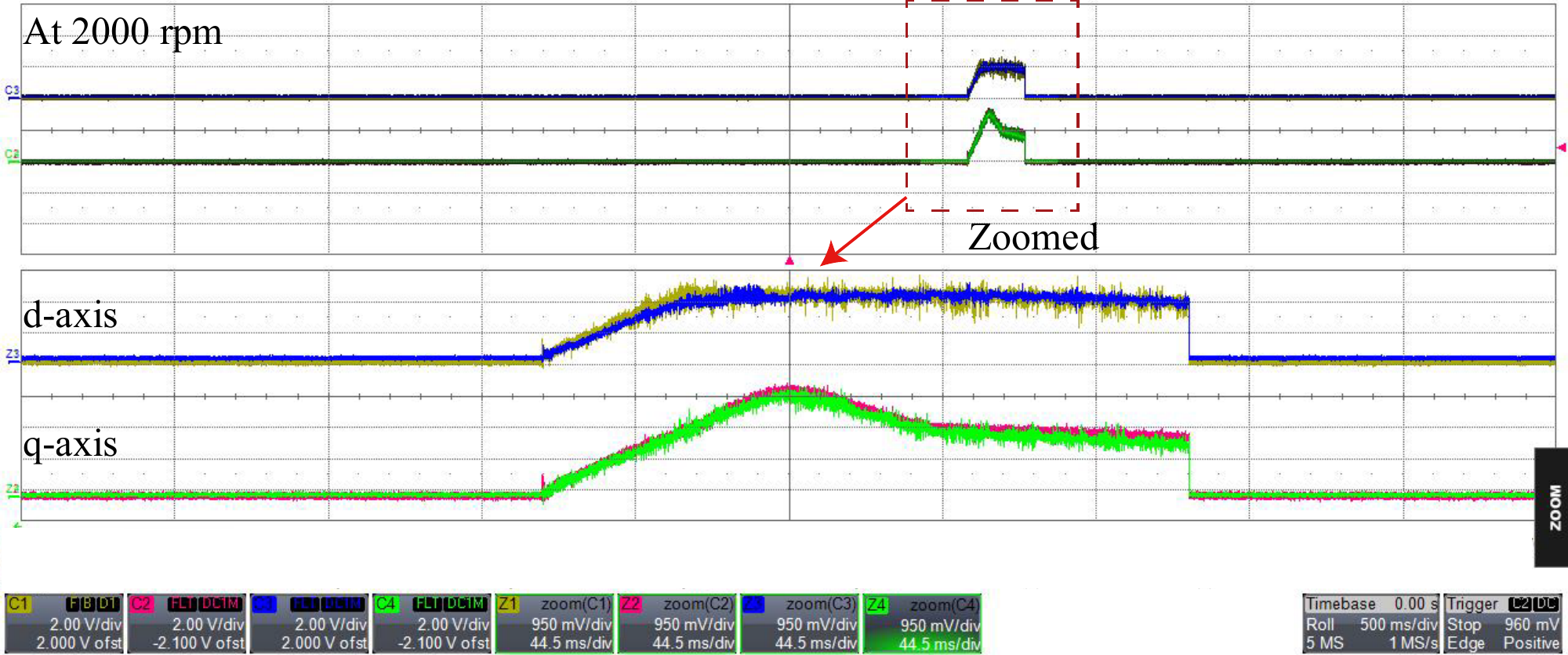} \label{fig8a}}
\caption{Experimental result for motor current response at varied speed employing proposed current control solution.}
\label{fig}
\end{figure}

\begin{figure}
\centering
\subfloat[PI current controller.] {\includegraphics[width=9cm]{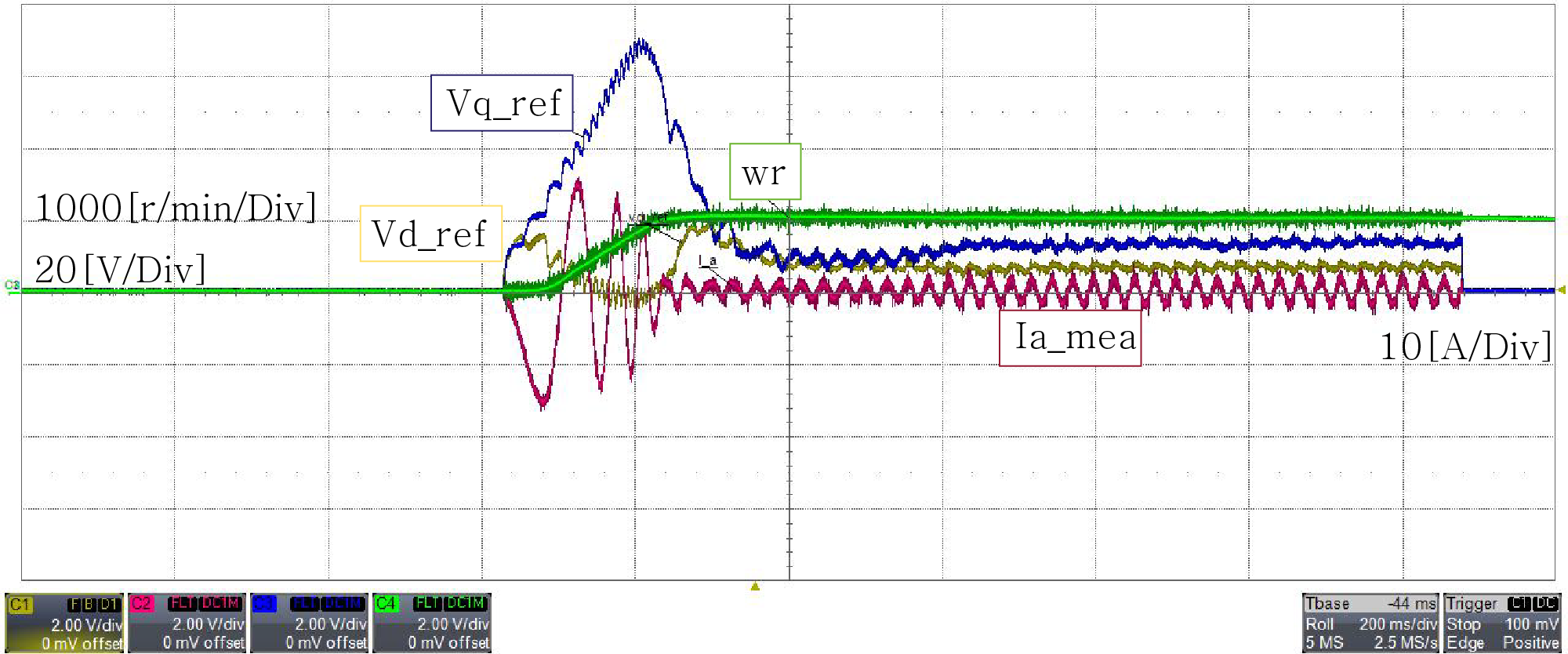} \label{fig7}} \\
\subfloat[DSR-based current controller.]{\includegraphics[width=9cm]{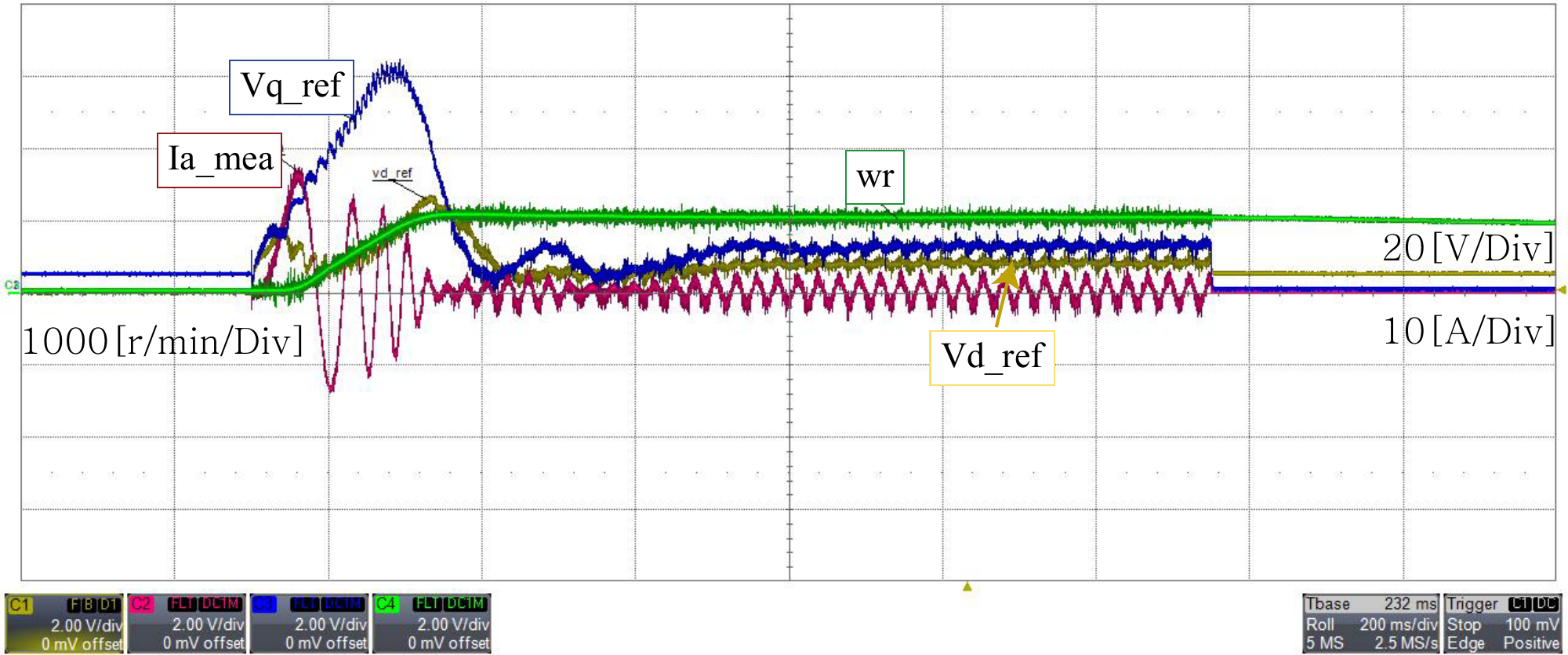} \label{fig8}}
\caption{Experimental result of motor response employing proposed and conventional current control solution.}
\label{figvd}
\end{figure}

\begin{figure}[htbp]
	\centerline{\includegraphics [width=9cm]{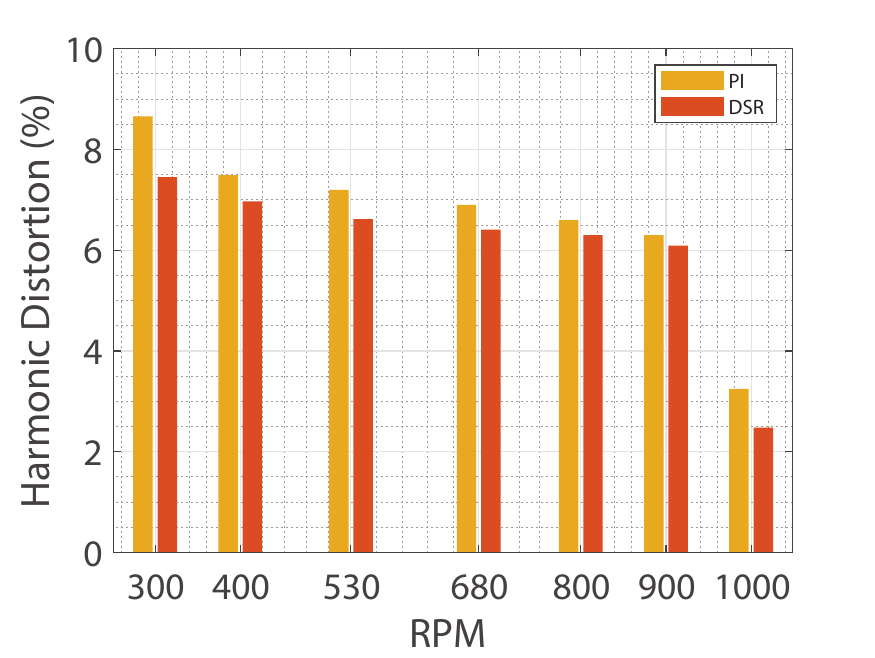}}
	\caption{Harmonic distortion comparative analysis based on simulation.}
	\label{figthd}
\end{figure}

To verify the efficacy of the proposed control solution, an experimental setup of an induction machine connected to a 15kw voltage source inverter (VSI) was implemented as depicted in Figure \ref{fig4}. Tms320f28388d from Texas Instruments is utilized to implement a spindle motor driver. The power rating for IM is $3.7$kw and the rated speed is 2000rpm. The control parameters used in simulation and experiment are shown in Table \ref{tab1}. 

Figure \ref{fig5} illustrates the simulation results of the traditional and proposed current control solution under varied speed and load conditions. The results demonstrate the effectiveness of the proposed control solution. The performance of the proposed control solution is comparable to the traditional control approach while exhibiting independence from model parameters and robust stability under load variation. Figure \ref{fig6a}-\ref{fig8a} shows a comprehensive view of the convergence between reference and measured currents at two distinct operational speeds(500rpm \& 2000rpm). The results depict the motor's dynamic response at various operational speeds and show good current convergence. Furthermore, the results illustrate the proposed nonlinear control law effectively responds to various speed references with dynamic current convergence. The overall closed-loop control performance is excellent with the proposed strategy.

 In Figure \ref{fig7}, we depict the reference voltages $(V_{dq}^*)$ and the motor response using state-of-the-art PI current control. whereas, Figure \ref{fig8} presents the reference voltage achieved through the proposed control solution, 
 along with the motor drive's performance. The results indicate that the proposed control strategy yields stable motor performance while utilizing less energy. This efficiency in energy consumption, coupled with the precise control of the motor, offers significant benefits for applications where conserving energy and maintaining precise motor operation are critical. The test results revealed excellent precision in tracking reference currents with minimal errors and the ability to achieve genuine independent control of flux and torque. In contrast to conventional methods, the proposed control approach doesn't require any tuning parameter, independent of motor parameters and is capable of exploring and showing superior dynamic performance. 

\begin{figure}
\centering
\subfloat[PI current controller.] {\includegraphics[width=7cm]{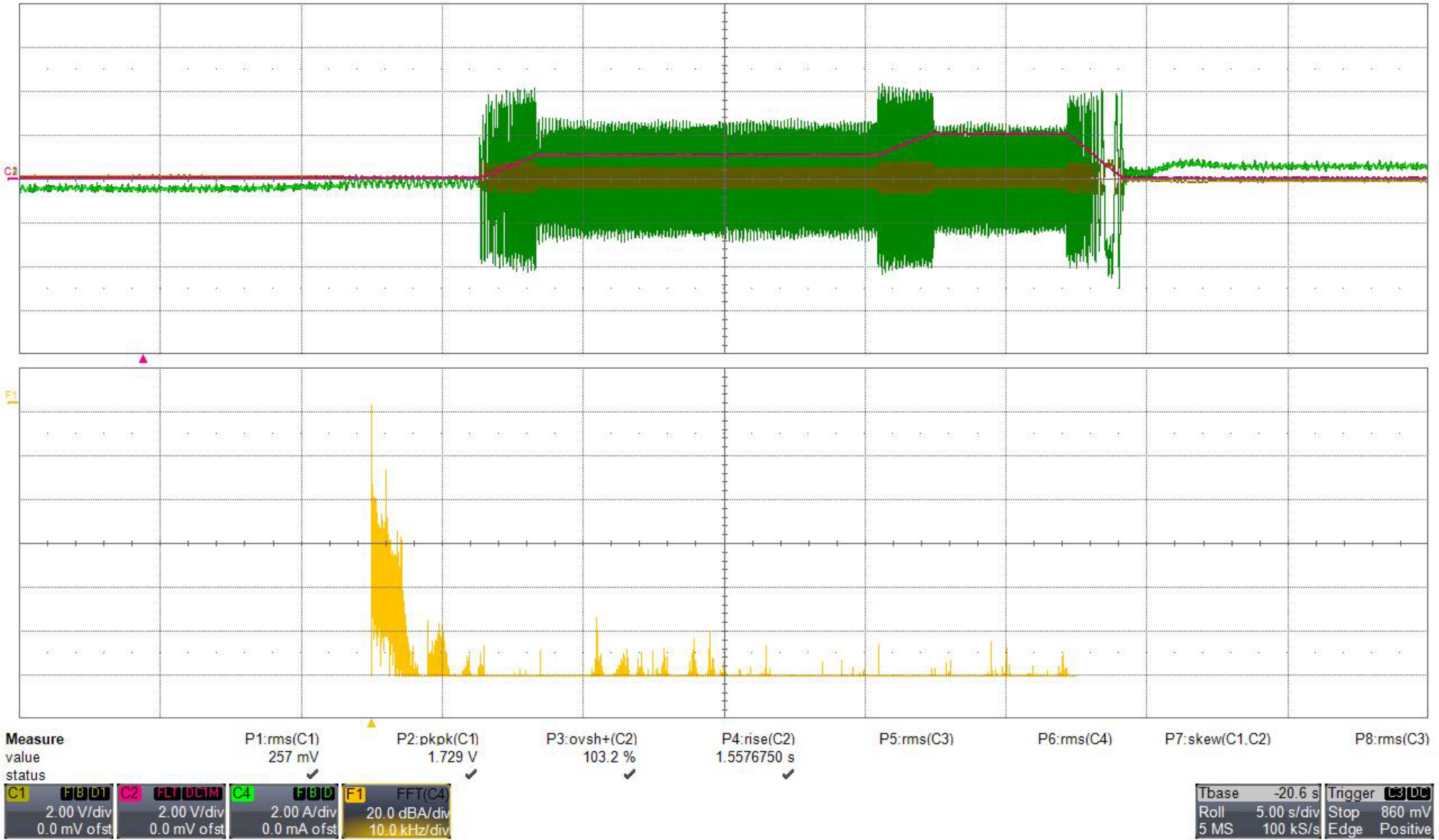} \label{fig_pi_thd}} \\
\subfloat[DSR-based current controller.]{\includegraphics[width=7cm]{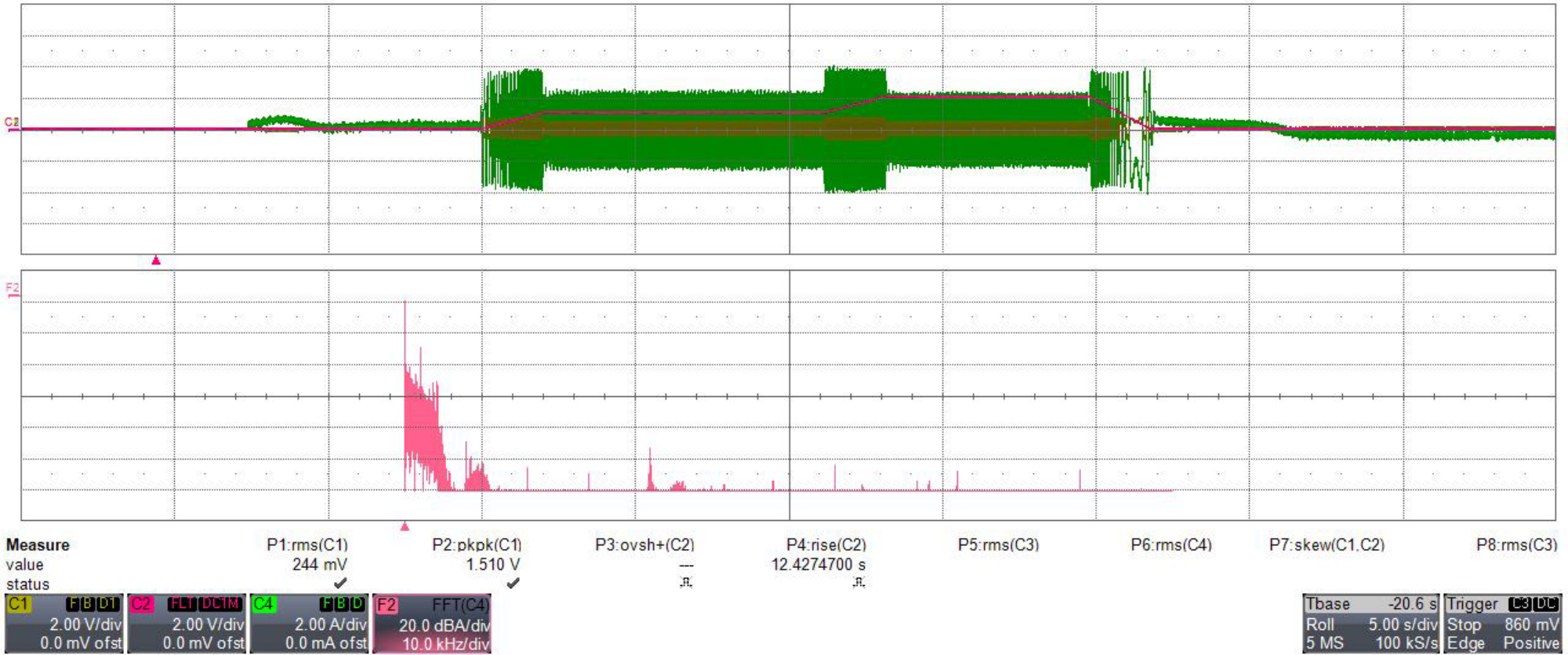} \label{fig_dsr_thd}}
\caption{FFT analysis of motor current employing traditional and proposed control solution.}
\label{figthdexp}
\end{figure}

Figure \ref{figthd} illustrates the harmonic distortion of the proposed and traditional current control solutions under various simulated speed references. Whereas the Figure \ref{figthdexp} shows experimental fft analysis. The result indicates that the proposed method consistently produces less harmonic distortion than the standard approach. This shows that the proposed control technique effectively reduces harmonic distortions, resulting in the system running cleaner and more efficiently. Reduced harmonic distortion indicates increased power quality and performance, making the proposed technique more suitable for real applications where harmonic distortion minimization is critical.

\begin{figure}
\centering
\subfloat[PI current controller.] {\includegraphics[width=7cm]{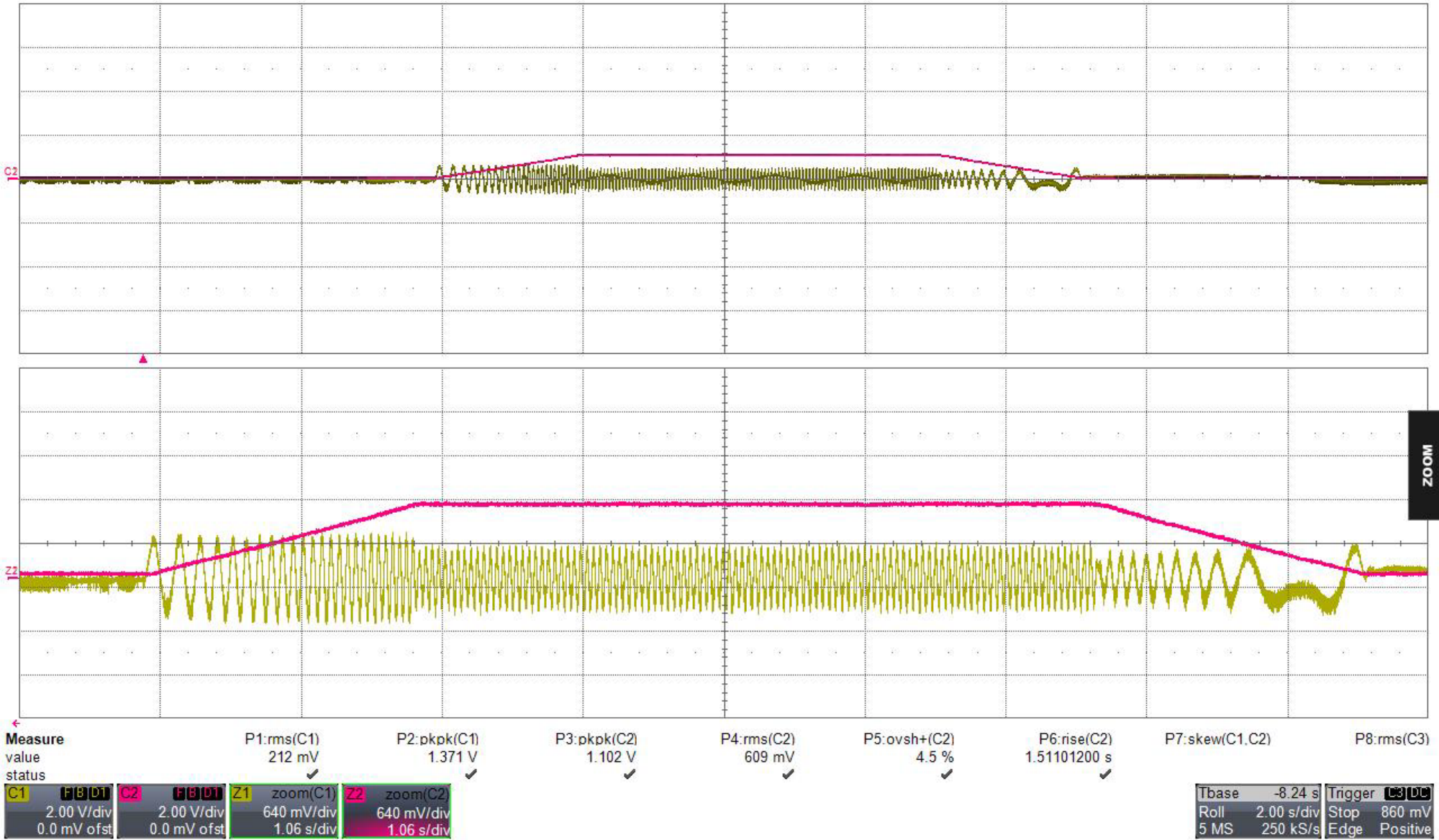} \label{fig_pi_acc}} \\
\subfloat[DSR-based current controller.]{\includegraphics[width=7cm]{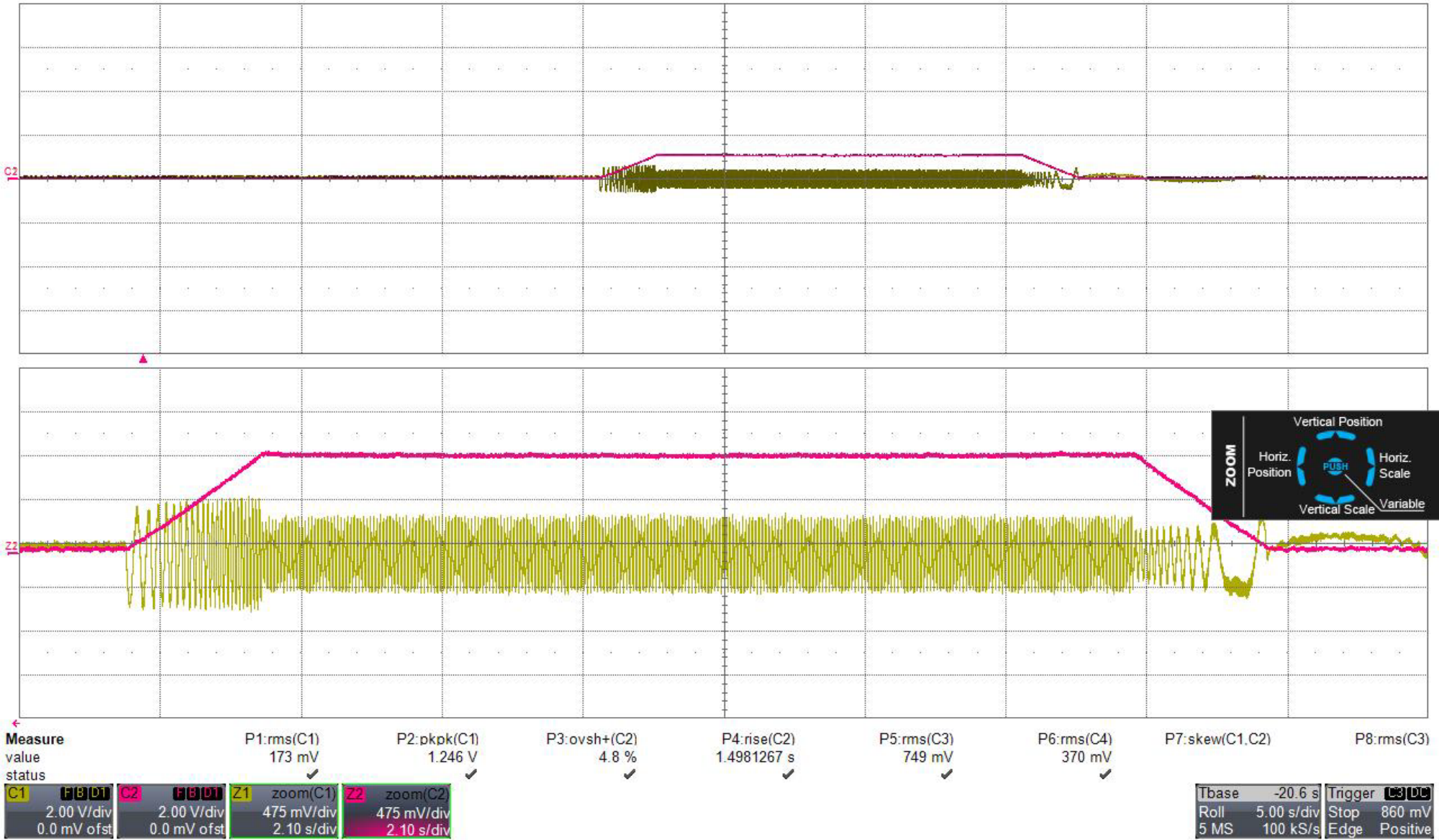} \label{fig_dsr_acc}}
\caption{Experimental results for motor behaviour during acceleration and deceleration of reference speed.}
\label{figacc}
\end{figure}

\begin{figure*}
\centering
\subfloat[PI current controller.] {\includegraphics[width=8cm]{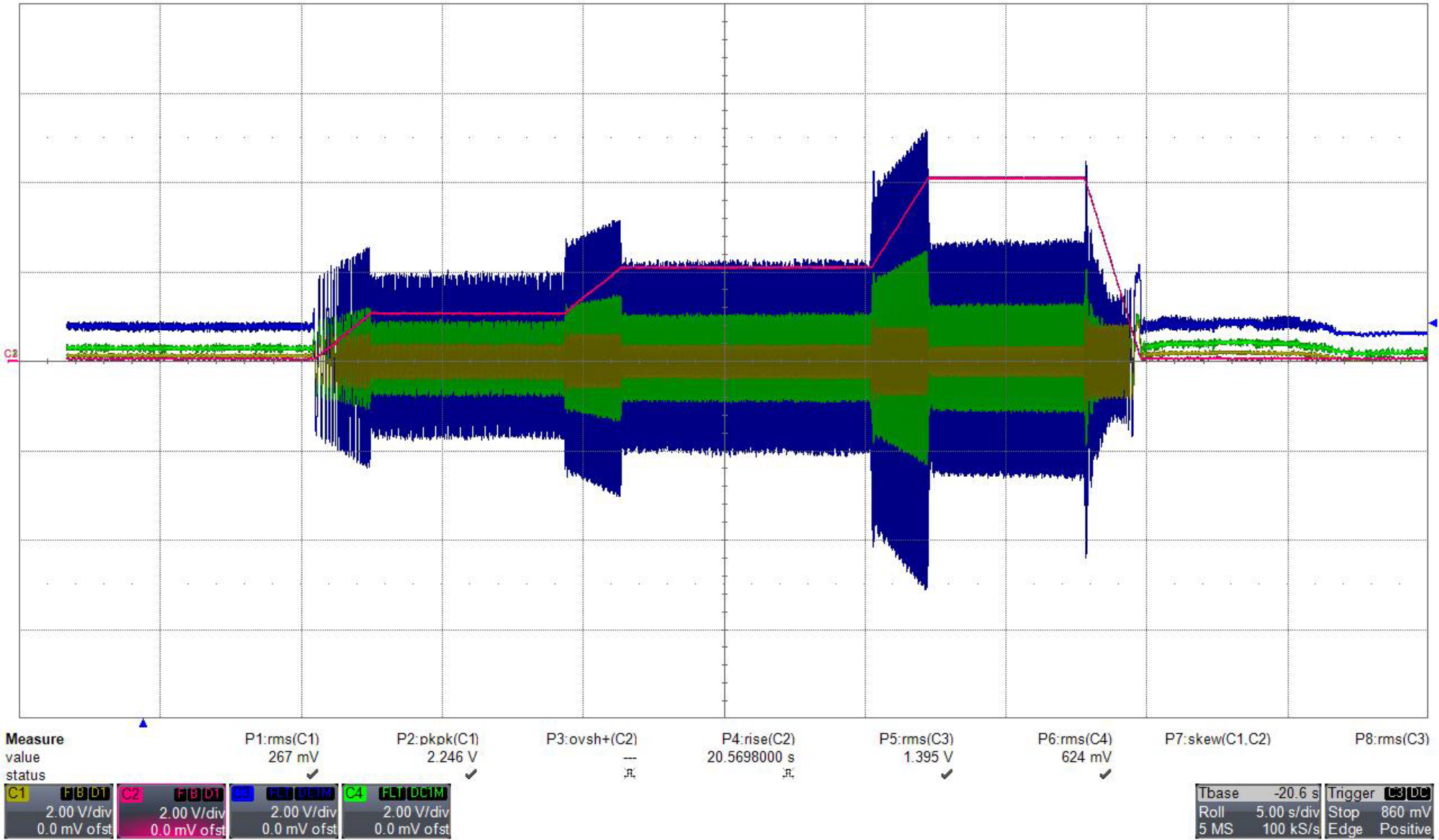} \label{fig_pi_res}} 
\subfloat[DSR-based current controller.]{\includegraphics[width=8cm]{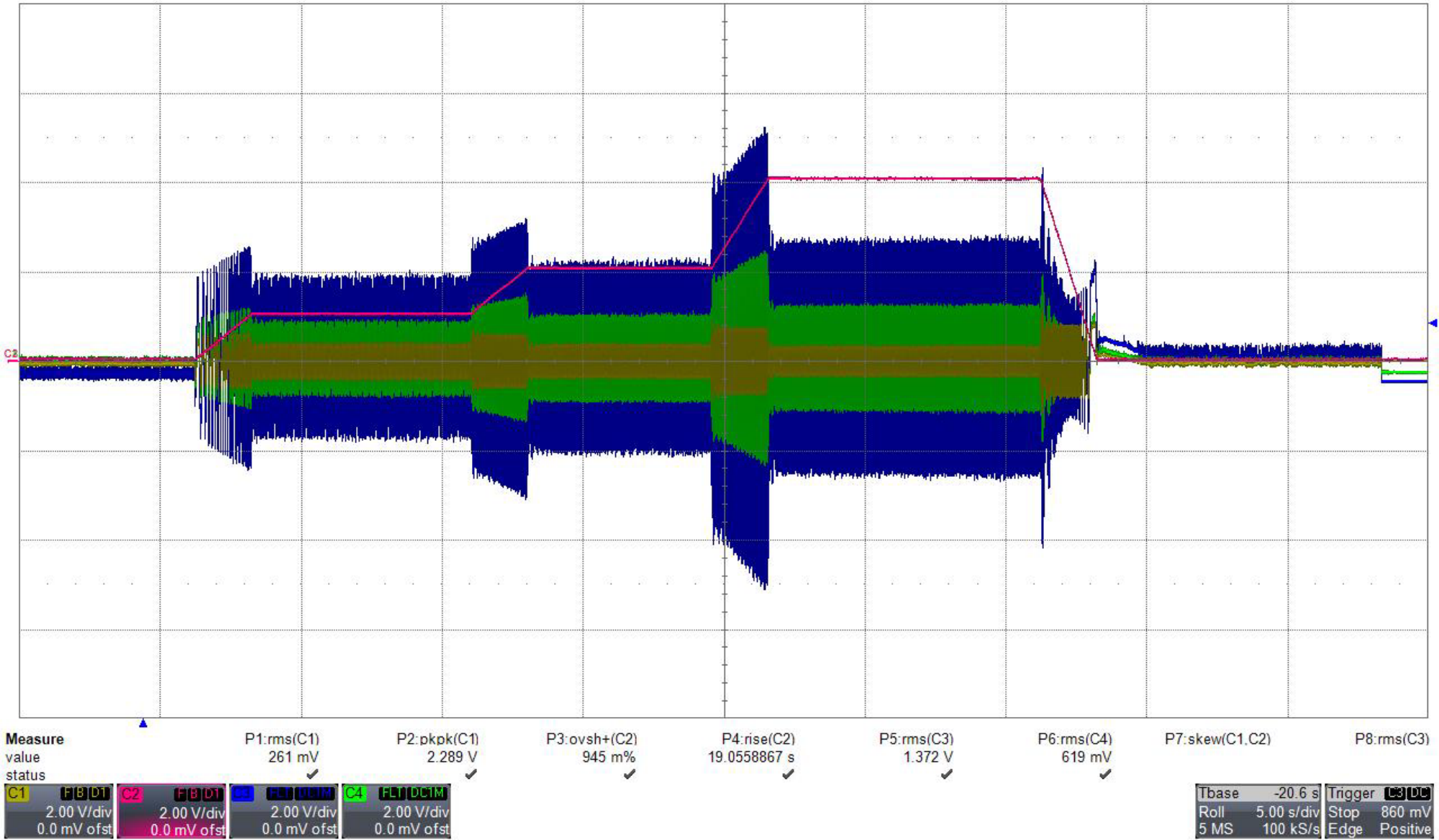} \label{fig_dsr_res}}
\caption{Experimental result of motor response employing proposed and conventional current control solution.}
\label{figexp}
\end{figure*}

Figure \ref{figacc} depicts the closed-loop control performance for acceleration and deceleration utilizing the proposed and PI-based current controller. The rms and peak-peak values of the current signal are minimal as compared to the PI-based inner current controller.  While maintaining good drive performance the proposed DSR-based current controller helps to optimize the efficiency, reliability and robustness. Figure \ref{figexp} shows the overall drive response at varied speed conditions. The results illustrate the excellent closed-loop performance and ability to track various reference inputs with the effectiveness of provided decoupled torque and flux control.


\section {Conclusion}
This paper introduces a DSR-based current control solution for Induction machines. Unlike traditional control scheme, the proposed control solution has ability to effectively decoupled electrical and mechanical dynamics and guarantee excellent performance of an ac drive for an induction machine. The proposed control solution provides a reduction in computational complexity compared to conventional AI-based control designs and offers ease of implementation in industrial applications. The proposed control approach provides excellent dynamic performance and improved system reliability. Leveraging AI strategy, provides good adaptability and response to sudden changes in operating conditions. Additionally, achieving true decoupling of torque and flux control enhances system flexibility and performance optimization. Comprehensive performance evaluation and experimental findings are provided to verify the performance of the proposed current control solution. 
\section*{Acknowledgment}

Development of Servo System Technology with a Current Response of 6.2 kHz and Power Regeneration for Automated Manufacturing Equipment Application Research project number: 20017351 Funded by Ministry of Trade, Industry and Energy (MOTIE) South Korea


\vspace{12pt}
\color{red}

\end{document}